\pdfoutput=1
\documentclass[12pt,preprint]{aastex}
\usepackage{subfig}
\shortauthors{Walkowicz et al.}  
\shorttitle{Kepler Flares}
\begin{document}

\title{White$-$light flares on cool stars in the \textit{Kepler} Quarter 1 Data}

\author{
Lucianne M. Walkowicz\altaffilmark{1},
Gibor Basri\altaffilmark{1},
Natalie Batalha\altaffilmark{2}, Ronald L. Gilliland\altaffilmark{3}, Jon Jenkins\altaffilmark{2}, William J. Borucki\altaffilmark{2}, David Koch\altaffilmark{2}, Doug Caldwell\altaffilmark{2}, Andrea K. Dupree\altaffilmark{4}, David W. Latham\altaffilmark{4}, Soeren Meibom\altaffilmark{4}, Steve Howell\altaffilmark{5}, Timothy M. Brown\altaffilmark{6}, Steve Bryson\altaffilmark{2}
}

\altaffiltext{1}{Astronomy Department, University of California at Berkeley, 601 Campbell Hall, Berkeley, CA 94720}
\altaffiltext{2}{NASA Ames Research Center, Moffett Field, CA 94035}
\altaffiltext{3}{Space Telescope Science Institute, Baltimore, MD 21218}
\altaffiltext{4}{Harvard-Smithsonian Center for Astrophysics, Cambridge, MA 02138}
\altaffiltext{5}{National Optical Astronomy Observatory, Tucson, AZ 85719}
\altaffiltext{6}{Las Cumbres Observatory Global Telescope, Goleta, CA 93117}

\begin{abstract}
We present the results of a search for white light flares on the $\sim$23,000 cool dwarfs in the \textit{Kepler} Quarter 1 long cadence data. We have identified 373 flaring stars, some of which flare multiple times during the observation period. We calculate relative flare energies, flare rates and durations, and compare these with the quiescent photometric variability of our sample. We find that M dwarfs tend to flare more frequently but for shorter durations than K dwarfs, and that they emit more energy relative to their quiescent luminosity in a given flare than K dwarfs. Stars that are more photometrically variable in quiescence tend to emit relatively more energy during flares, but variability is only weakly correlated with flare frequency. We estimate distances for our sample of flare stars and find that the flaring fraction agrees well with other observations of flare statistics for stars within 300 pc above the Galactic Plane. These observations provide a more rounded view of stellar flares by sampling stars that have not been pre-selected by their activity, and are informative for understanding the influence of these flares on planetary habitability. 

\end{abstract}


\section{Introduction}

Flares are short-lived but intense releases of energy, caused by the reconnection of loops of magnetic field in the outer atmospheres of active stars. During these reconnection events, the field geometry reconfigures to a lower energy state, often accelerating energetic particles down towards the stellar atmosphere and causing a plethora of observable effects over a wide range in wavelength \citep[for a recent review of flare processes, see][]{benz2010}. On the Sun, our ability to spatially resolve the surface allows flares to be observed in rich detail with high cadence, and so much of our understanding of flare processes and their consequences comes from observations of solar flares. Unfortunately, the luxury of spatial resolution is not available for flares on stars other than our Sun, and so stellar flare studies rely instead on time$-$resolved photometry or spectroscopy. 

Stellar flares long ago revealed themselves to be somewhat different from the standard solar model. Stars cooler than our Sun, such as the K and M dwarfs, have flares that seem both surprisingly energetic (flares on active M dwarfs are typically 10 $-$ 1000 times as energetic as solar flares) and qualitatively different than solar flares, showing strong  continuum or ``white light'' emission, which resembles a 9000$-$10,000K blackbody superimposed over the quiet spectrum of the star \citep{haisch1991,hawley1991}. White light emission in solar flares is usually concentrated in small areas near the footpoints of emerging magnetic field lines, is spatially associated with hard X$-$ray emission, and does not appreciably affect the integrated optical brightness of the Sun \citep[e.g.][]{watanabe2010}. By contrast, dramatic brightening due to white light emission is the photometric hallmark of cool star flares. The white light emission in these flares has been inferred to cover a much larger fraction of the visible stellar surface than the white light emission seen in solar flares-- for example, the recently reported white light flare emission on the M dwarf YZ CMi was inferred to cover 0.22$\%$ of the visible disc \citep{kowalski2010}, or roughly $\sim$400 Mm$^2$, while the 2002 solar flare discussed in \citet{potts2010} had a $\sim$20 Mm$^2$ emitting area. These observed differences likely have to do with the changing topology of the magnetic field as one moves to cooler stars. Active stars near the convective boundary (M$_{\star}$$<$0.4M$_{\Sun}$) tend to have large scale, non-axisymmetric fields and larger areal coverage of active regions \citep{saar1985,cjk1996,donati2008,reiners2009}. Although stellar flares are fundamentally multiwavelength phenomena and have been observed across the electromagnetic spectrum \citep[e.g.][]{hawley2003,osten2005,hawley2007} photometric flare campaigns in the optical tend to focus on observations in the blue due to the higher contrast of the flare white light emission with the photosphere (and thus enhanced detectability of the flare). 

Open questions regarding the overall rate and properties of flares have recently come to the forefront, due to the effects of flares on planetary habitability and as a source of Galactic transients in new time domain surveys \citep{segura2010,tyson2006}. Typically, stellar flare research has focused on detailed studies of a few of the most active known stars, largely due to the reluctance of astronomers and time allocation committees to spend telescope time observing a star which may not do anything in the course of a night. These studies have concentrated in particular on the most active M dwarfs, the dMe stars-- so dubbed due to their powerful H$\alpha$ (e)mission lines, indicative of strong magnetic activity. The seminal work of \citet{lme1976} determined the rates and energy distributions of flares for 8 active M dwarfs, and has long been the most detailed optical study of flare rates to date. However, little is known about how often flares occur on stars with moderate activity. Cool stars with H$\alpha$ {\em absorption} are often also chromospherically active, albeit at a lower level, and have also been serendipitously observed to flare amongst stars {\em not} pre-selected for their activity \citep{walkowicz2008,kowalski2009}. In both the Sun and dMe stars, the probability distributions of flare energies are best described by a power law, indicating that small flares are more common than large  \citep{charbonneau2001,shakhovskaia1989,lu1991}. Much work has been done on the Sun to quantify the index that sets the slope of this power law, as the proportion of small flares to large determines whether so-called ``nanoflares'' are capable of being a significant source of coronal heating. However, dMe stars are quite different from the Sun, and it is not yet known whether the distribution of flare energies can be described by a power law for stars of all spectral types, and whether the indices of these power laws are the same (indicating a common physical process). 

In this paper, we investigate flares found from observations of stars by the \textit{Kepler} mission \citep{keplermain}. While \textit{Kepler}'s primary objective is the detection of exoplanets, its constant monitoring of thousands of cool stars allows us to probe flares in a sample unlike any targeted for flare searches before. \textit{Kepler} provides unmatched precision photometry, but does so only in a broad bandpass filter (400 $-$ 900 nm). The contrast between the flare emission and the quiescent stellar photosphere is therefore lower than for observations performed in a blue filter, and the lack of color information makes it difficult to use previously defined metrics such as flare color indices \citep{kowalski2009}. We therefore searched for the most obvious deviations from quiescence, likely to be astrophysical in nature rather than instrumental, and report our first results from a search for white light flares in the \textit{Kepler} Quarter 1 observations here. In the next section, we outline the sample selection and the flare search algorithm, as well as how each flare was measured, including a new definition for a useful metric we use here. In the last section, we discuss the implications of our results and potential areas of future research. 

\section{Data Selection and Analysis}

The \textit{Kepler} Quarter 1 long cadence data consist of 33.5 days of photometry sampled every 30 minutes, taken between 13 May 2009 and June 15 2009. In this paper, we work from the raw simple aperture photometry provided in the publicly released Quarter 1 FITS files, under the AP$\_$RAW$\_$FLUX keyword. The ``raw'' photometry is not truly raw: it has been corrected for the background, flat fielded, subjected to cosmic ray removal and aperture summed. The resulting photometry still contains some instrumental effects and artifacts, most notably long term trends over the quarter due to differential velocity aberration \citep{keplerinstrument,keplerlongcadence} We remove these trends and instrumental effects with our own analysis pipeline, described in detail in \citet{basri2010b}. In brief, our treatment of the data removes likely instrumental trends but leaves shorter-term spot-like variability intact. While some of the long term trends we remove over Q1 may be astrophysical in nature, we err on the side of caution as we cannot distinguish between true and instrumental trends over only a single month of data. For the purposes of searching for transient events like flares, long trends in the data are immaterial, as they occur over a much longer timescale (i.e. weeks) than the relatively brief duration of flares (i.e. hours). 

\subsection{Sample Selection}

Our flare search sample consists of all stars that are classified in the \textit{Kepler} Input Catalogue \citep[KIC;][]{kic} as having logg $\ge$ 4.2 and T$_{eff}$ $\le$ 5150 K,  corresponding to 23,253 main sequence stars later than K0V\footnote{Our sample includes all exoplanetary search targets that are classified in the KIC, but excludes targets assigned to \textit{Kepler} GO observers.}. The entire sample is detailed in Table 1, which is available in its entirety in the electronic version of this paper. Candidate flare events were initially identified by an automated search of all cool dwarf lightcurves meeting the KIC criteria outlined above. Our search algorithm has three adjustable parameters, the threshold, smoothing, and width. The threshold is the multiple of the standard deviation, excluding points that are extreme outliers from the norm, above which a point is flagged as a significant brightness change. The smoothing is the number of points over which to median filter the lightcurve to minimize point-to-point variations-- the smoothed lightcurve is ultimately subtracted from the raw lightcurve, which removes most of the quiescent stellar variability and allows shorter timescale events to be found more easily. Lastly, the width is the minimum number of contiguous points that must lie above the threshold before an event will be flagged as a potential flare. The appropriate values for these search parameters were developed by experimentation between the automatic flare finder and a by-eye flare search on a test set of stars, the M dwarfs. A subset of the $\sim$2300 M dwarf lightcurves was first searched by eye for all outliers that appeared to be flares. These same lightcurves were then searched automatically using different values for the threshold, smoothing and width until the search algorithm was capable of finding the same set of flaring stars as the human eye. Naturally these two sets were not exactly the same, as the automatic search more easily found small, marginal candidate events than the human eye was capable of finding. We did ascertain that anything that was obviously a flare to the eye (possessing a clear impulsive rise and exponential decay) was found by the automatic flare finder. Ultimately we chose to smooth the lightcurves over a 10 hour interval, and events were flagged as candidate flares when a minimum of three contiguous points were found above a threshold of 4.5 times the standard deviation. As the \textit{Kepler} lightcurves are sampled every 30 minutes, requiring a minimum number of points biases our search towards flares with a duration of 1.5 hours or longer, but this requirement also helps to distinguish astrophysical events from instrumental artifacts. 
 
A search with the above parameters yielded 5784 candidate events, which were then vetted by eye and culled further to 2741 possible candidate flare stars. These 2741 candidate flare stars were again vetted by eye, and 373 stars were identified as having obvious flares showing the classical impulsive rise / exponential decay lightcurve shape. An additional 565 stars were identified as marginal cases, having potential-but-not-certain flares. In this paper, we focus our analysis on the 373 certain flare stars identified. We intend to revisit smaller events in the future, preferably when pixel-level data is available to rule out particle hits and other detector effects. 

\subsection{Measurement of Flares}
The \textit{Kepler} long cadence data is formed on board the spacecraft by summing 270 6-second integrations into successive blocks of 29.42 minutes. As such it is not {\em ideally} suited to the study of flares, as most flares evolve on timescales faster than the $\sim$30 minute cadence is capable of resolving. The peak of each flare is integrated into one cadence or another, but has somewhat lower contrast than if the cadence were shorter. For each lightcurve with flares, we masked out points tagged as flares and median filtered the remaining lightcurve to create a lightcurve with quiescent variability but no flares. We then subtract this quiescent lightcurve to remove slower modulation due to the presence of spots on the quiescent star. We then subtract off this fit, run the flare finder again to flag contiguous points that deviate above our threshold, and integrate the points that are tagged as part of the flare. What results is essentially a photometric equivalent width for the flare, or
$$
EW_{phot} = \int\frac{F_f - F_q}{F_q} dt
$$

where $F_f$ and $F_q$ denote the flaring and quiescent flux, respectively. By analogy to equivalent width this quantity has the units of time, but can be intuitively thought of as the time interval over which the quiescent star emits as much energy as was released during the duration of the flare. EW$_{phot}$ is a differential quantity, independent of distance and measured relative to the quiescent star. 

We also calculate the duration of each flare, record the value of the brightest point in the flare, and if multiple flares occur on a single lightcurve, the time between each flare. In the interest of working with the data as is, we chose not to integrate assuming a typical flare lightcurve peak+exponential decay shape, nor do we perform detailed modeling of individual flares in this work. These studies are better suited to short cadence data (which was unavailable for the stars identified in this paper) and will be visited in future complementary work. 

\section{Results}

\begin{figure}
\centering
\subfloat[Example lightcurves of four flaring M dwarfs.]{
\includegraphics[width=0.75\textwidth]{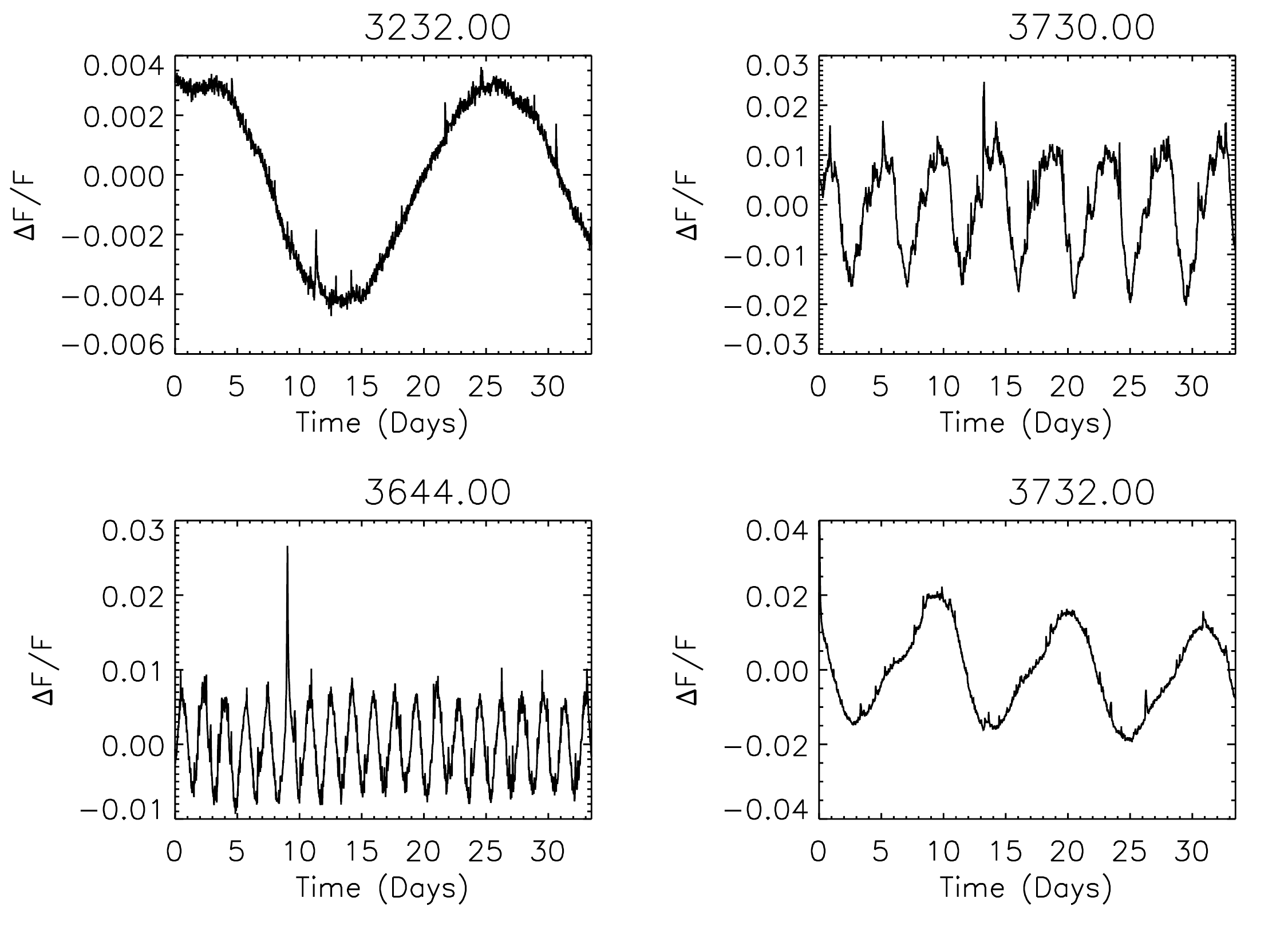}
}\\
\subfloat[Example lightcurves of four flaring K dwarfs.]{
\includegraphics[width=0.75\textwidth]{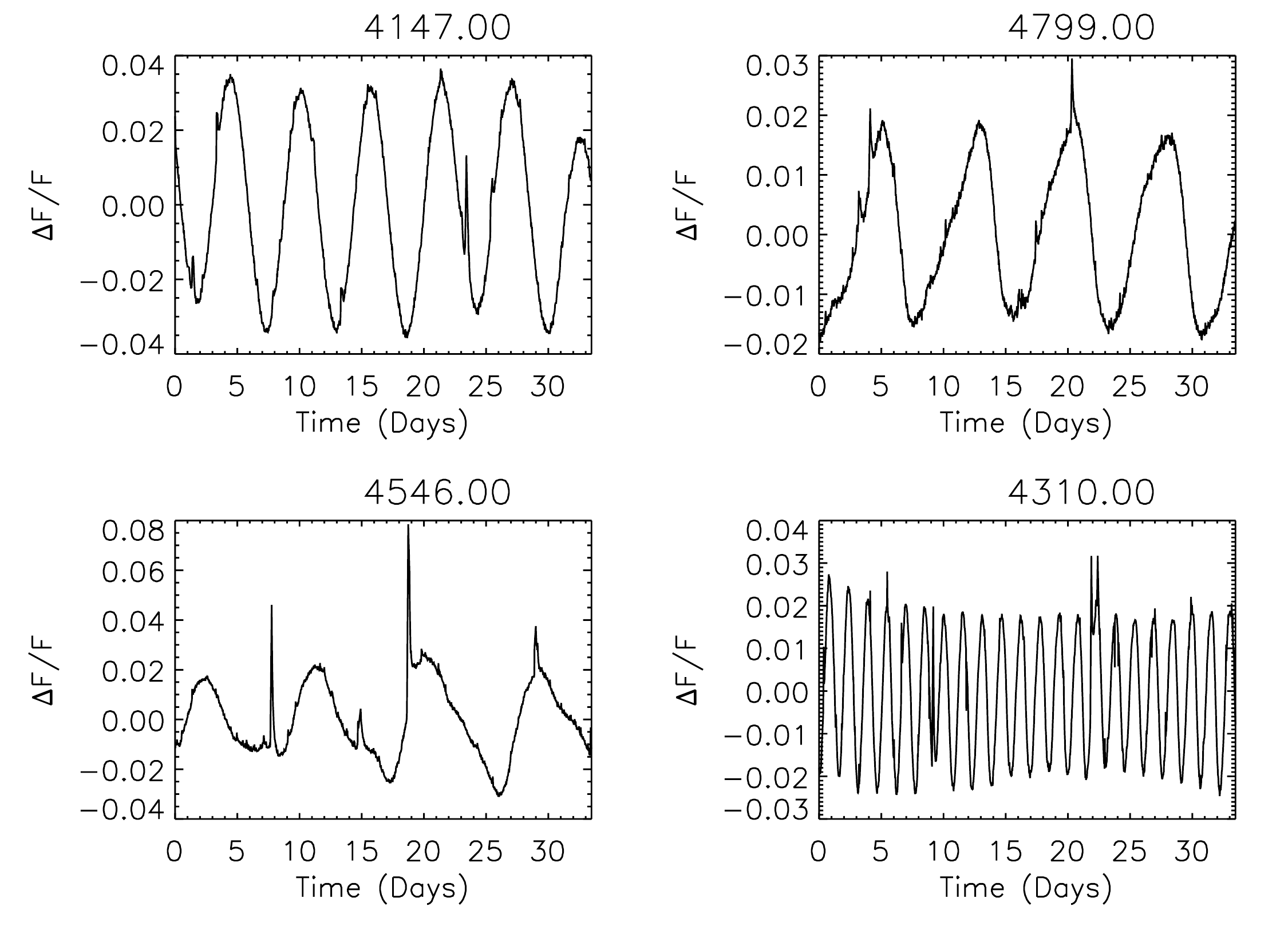}
}
\caption[]{Four example flaring M dwarf (top) and K dwarf (bottom) lightcurves from Quarter 1. The effective temperature of the star is listed above each plot.}
\label{flare_lcexample_mk}
\end{figure}

\begin{figure}[ht]
\centering
\subfloat[Detail of subtracted lightcurve for an example flaring M dwarf.]{
\includegraphics[width=0.75\textwidth]{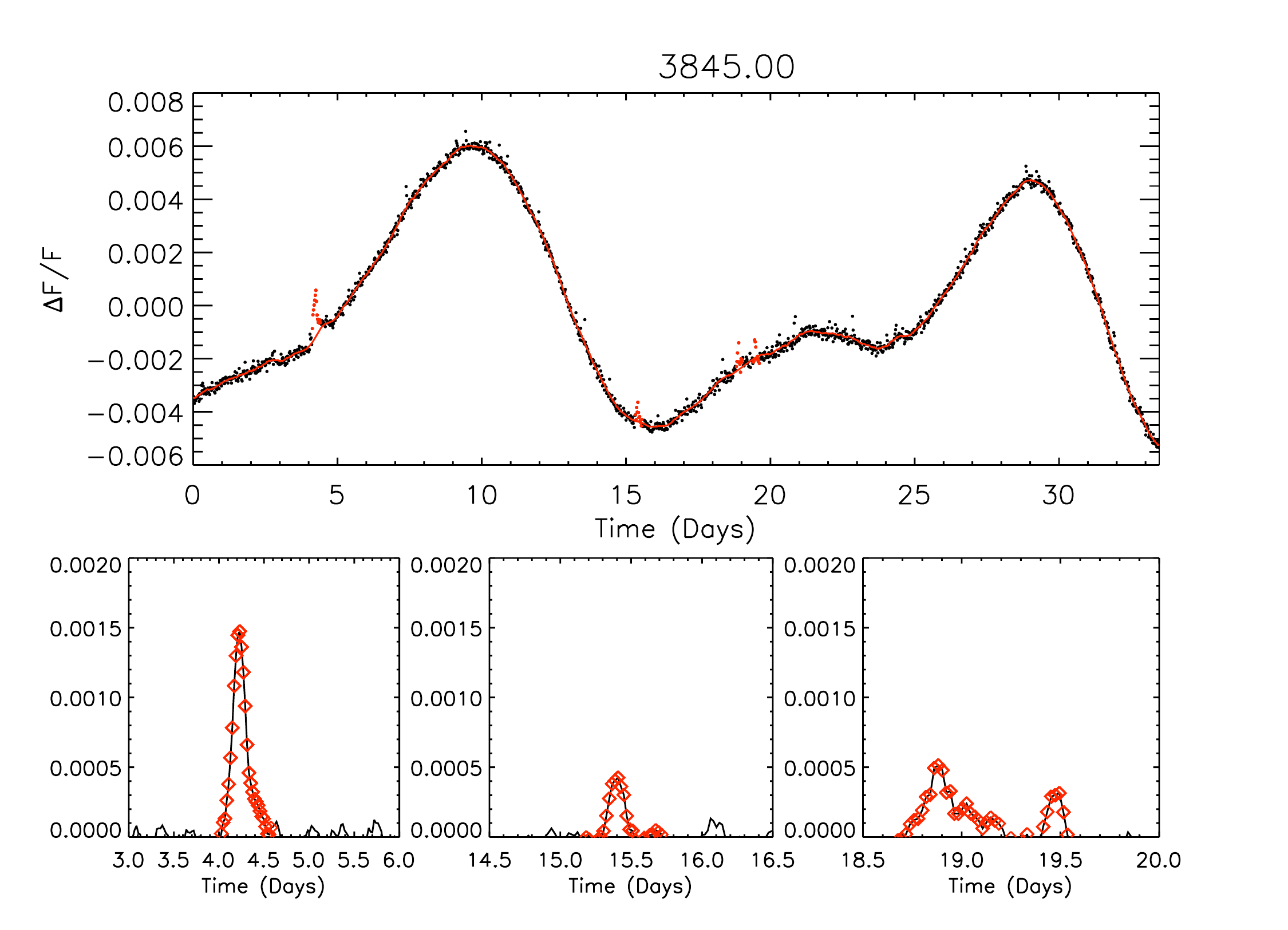}
\label{fig:subfig1}
}\\
\subfloat[Detail of subtracted lightcurve for an example flaring K dwarf.]{
\includegraphics[width=0.75\textwidth]{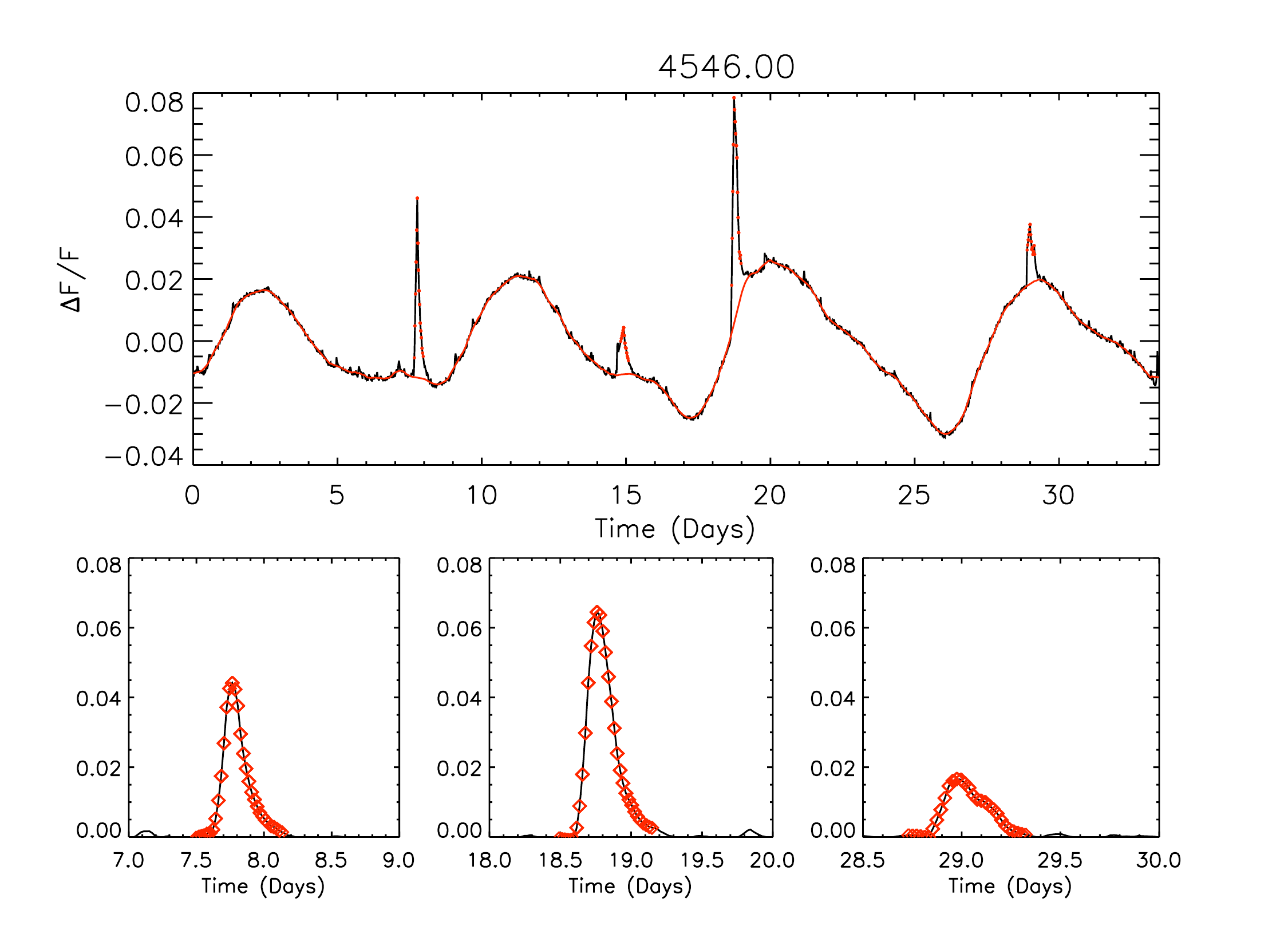}
\label{fig:subfig2}
}
\caption[Optional caption for list of figures]{Two example lightcurves for one of the flaring M dwarfs (top) and K dwarfs (bottom). In the top panel of each subfigure, the full lightcurve is shown, with the ``continuum'', or fit to the quiescent variability, overplotted in a solid red line. Points which meet the flare criteria outline in the previous section are marked with red points overlying the data. In the bottom panels, three example flares from each lightcurve are shown at an expanded scale, where the quiescent stellar variability has been subtracted.}
\label{flare_lcexample_mk_detail}
\end{figure}

The \textit{Kepler} data offers a truly remarkable opportunity both to capture flares and to observe the quiescent variability of the star. Figure \ref{flare_lcexample_mk} shows four examples of flares found on M and K dwarfs, respectively, with the KIC effective temperatures shown at the top of each panel. In Figure \ref{flare_lcexample_mk_detail}, we show two of these lightcurves in further detail: the top panel shows the data as black points with flare points marked in red, and the fit to the star's quiescent variability is overplotted as a thick red line. Below, subpanels show examples of flares at an expanded scale, with the quiescent stellar flux removed$-$ red diamonds indicate the points flagged as belonging to the flare event.  

Table 2 presents the resulting number of flares for each star, the median EW$_{phot}$, flare peak brightness, duration, time between flares, the flare frequency, and percentage of time spent flaring (this table is available in its entirety in the electronic version of the paper).  

\begin{figure}
\begin{center}
\includegraphics[width=0.95\textwidth]{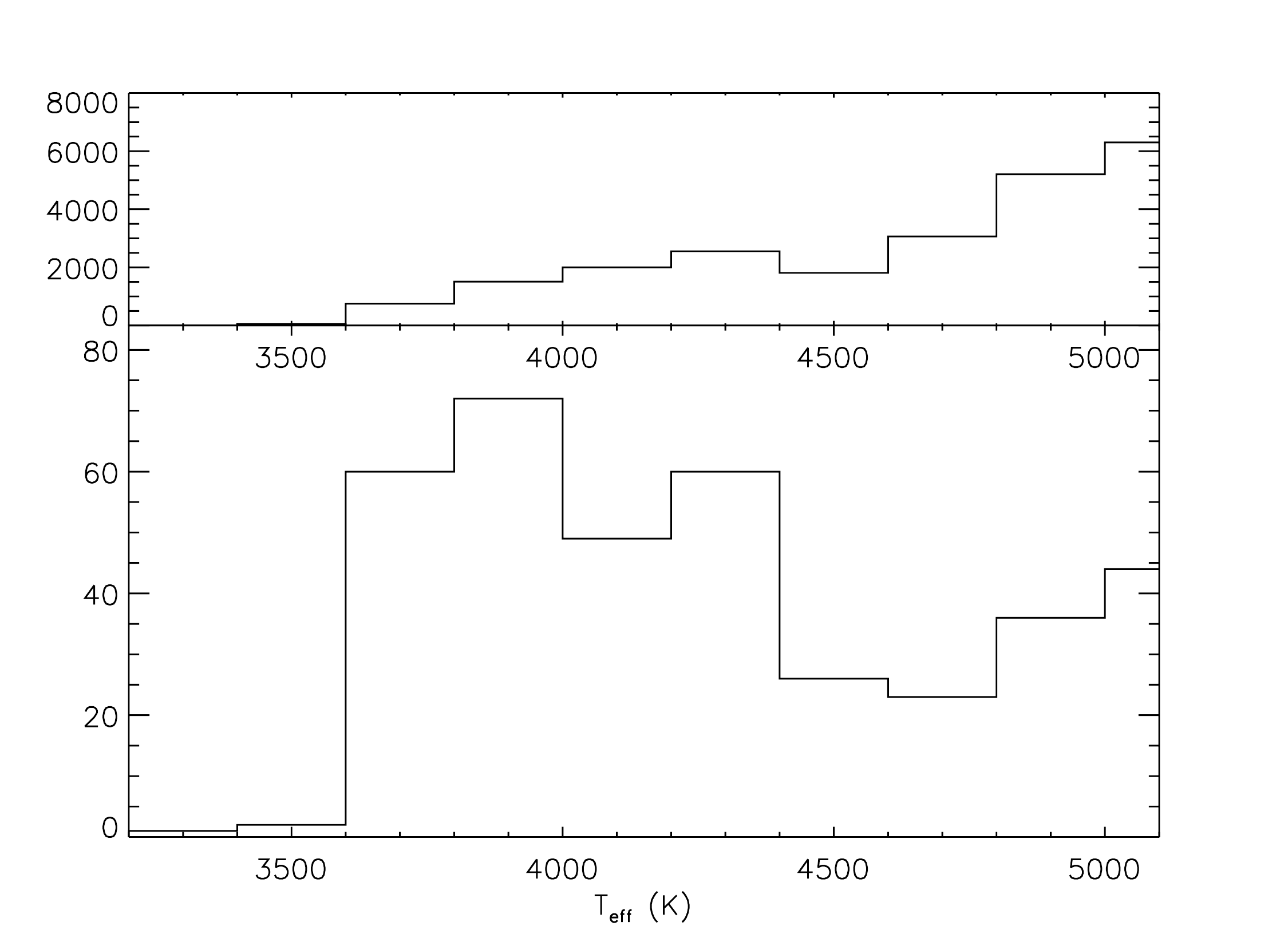}
\end{center}
\caption{In the upper panel,we show a histogram of the effective temperature distribution of the entire cool dwarf sample, while the effective temperature distribution of the flare stars is shown below. The number of stars found to flare increases with decreasing effective temperature, partly due to the increasing contrast of white light flares.}
\label{flare_flrfracteff}
\end{figure}

In the top panel of Figure \ref{flare_flrfracteff}, we show the histogram of the effective temperature distribution of the entire cool dwarf sample, while the T$_{eff}$ distribution for the sample of flaring stars appears in the lower, larger panel. Clearly, the number of stars on which flares were found increases with decreasing effective temperature, caused (at least in part) by the increasingly high contrast of white light flares for cooler stars. 

\begin{figure}
\begin{center}
\includegraphics[width=0.95\textwidth]{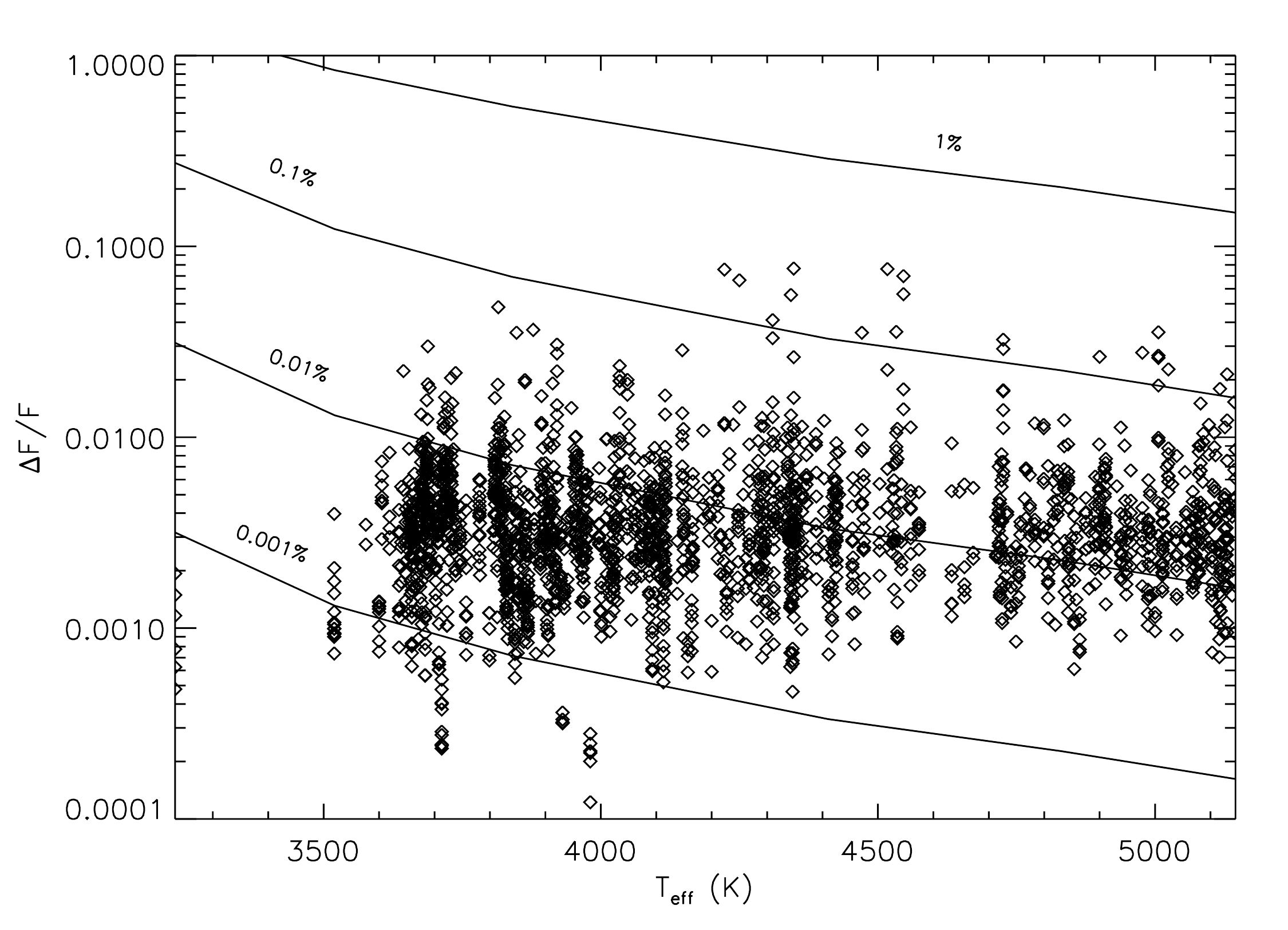}
\end{center}
\caption{Maximum brightness enhancement for all flares observed versus effective temperature. The data are overplotted with lines of the expected brightness change for a 10,000 K blackbody superimposed on the star at various filling factors: 0.001$\%$, 0.01$\%$, 0.1$\%$ and 1$\%$ (see the related text for a more complete explanation). The distribution of peak brightness change is more or less flat, as it is possible to detect smaller flares on the M dwarfs due to enhanced contrast between the flare emission and photosphere.}
\label{flare_peakvteff}
\end{figure}

In order to better quantify the effect of the flare-to-photosphere contrast on our flare detections, we added a 10,000K blackbody curve at varying areal coverage (so-called ``fill factor'') to a series of cool stellar template spectra \citep{pickles} and convolved these synthetic flare spectra with the \textit{Kepler} filter response. We then computed the instantaneous change in brightness ($\Delta$F/F) we would expect to find for each spectral type for different fill factors. In Figure \ref{flare_peakvteff}, we show effective temperature versus the peak $\Delta$F/F in each flare, overplotted with lines of constant fill factor for a 10,000K blackbody. 

Figure \ref{flare_peakvteff} shows a somewhat flat distribution of flare brightness enhancement over the range of T$_{eff}$ in our sample. The relative dearth of very small flares with increasing effective temperature represents the detection threshold imposed by contrast effects$-$ white light flares have a lower contrast on warmer stars, therefore small flares are more likely to go undetected. As effective temperature decreases, we are able to detect more small flares. There are fewer large flares detected towards cooler stars, which might be expected if the K dwarfs behave similarly to observed trends on M dwarfs: SDSS Stripe 82 observations of flaring M dwarfs show that higher mass cool stars tend to have more luminous flares than their cooler, less massive counterparts (Kowalski et al. 2010b). While at first glance the absolute value of the flux enhancements might seem quite small, we remind the reader that the relatively broad \textit{Kepler} bandpass effectively ``washes out'' the effect of the flare on the photometry$-$ for reference, aforementioned megaflare on YZ CMi, a cool M4.5V star, was 6 magnitudes in U, but only a 0.22\% fill factor \citep{kowalski2010}. 

\begin{figure}
\begin{center}
\includegraphics[width=0.95\textwidth]{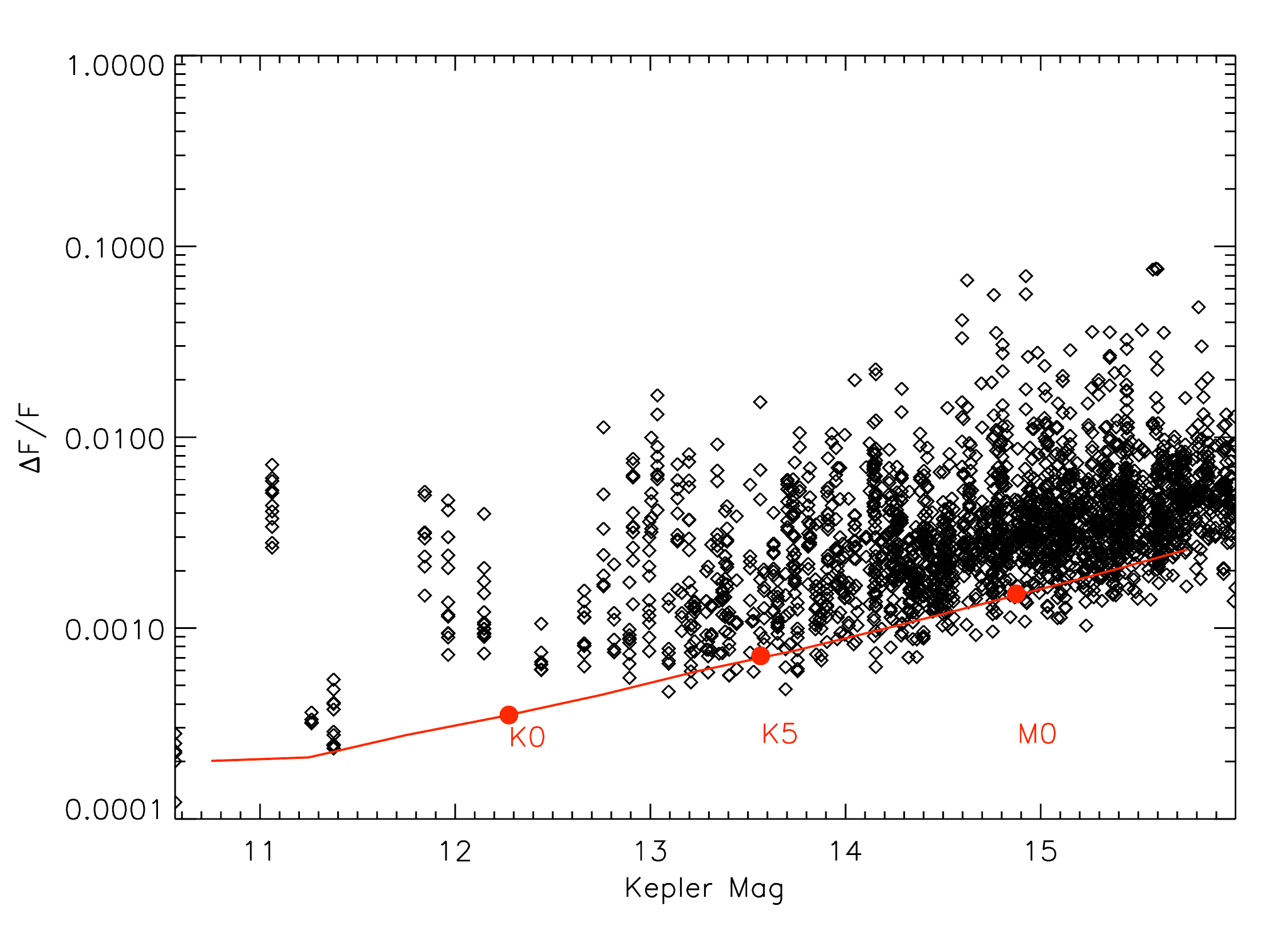}
\end{center}
\caption{Flare peak height as a function of Kepler magnitude, with our detection threshold overplotted (at least three consecutive points in the lightcurve must lie above this threshold to be tagged as a potential flare). As three red circles, we show where  K0, K5 and M0 dwarfs would intersect this threshold if these stars were at a distance of 200 pc. At a given distance, intrinsically more luminous stars are brighter and therefore less noisy, such that the detection threshold is lower. As a result, although the emission from flares on hotter, more luminous stars has lower contrast with their photospheres, one can actually detect a smaller flare on them at a given distance because of their relatively lower noise.}
\label{flare_peakvkepmag}
\end{figure}

 The detectability of a flare is a function not only of the relative contrast of the white light emission against the photosphere, but also the relative height of the flare against the intrinsic noise in the quiescent lightcurve, which is a function of magnitude. Candidate flares are required to have three consecutive points above our threshold as described in the previous section. Figure \ref{flare_peakvkepmag} shows the flare peak heights as a function of Kepler magnitude, with our detection threshold overplotted. As three red filled circles, we show where K0, K5 and M0 dwarfs would intersect this threshold if they were placed at a distance   of 200 pc. At a given distance, intrinsically more luminous stars are brighter and therefore less noisy-- therefore, although the emission from flares on these stars has lower contrast with their photospheres, one can actually detect smaller flares on them compared with cooler stars {\em at a given distance} because of their relatively lower noise.

Assuming that flares are associated with the same magnetic field formations that produce starspots, one might also expect there to be a correlation between the photometric variability of the lightcurve due to spots and the presence of flares. We quantify the bulk variability as ``variability range'', the range between the 5th and 95th percentile amplitude in the lightcurve (discussed in further detail in Basri et al. 2010b). In the center panel of Figure \ref{flare_histrangevenergy}, we show the relationship between the median EW$_{phot}$ for the K dwarfs (black asterisks) and M dwarfs (red diamonds) and the photometric range in millimagnitudes. The flanking panels show the histograms of the range (at top) and EW$_{phot}$ (at right) for the K and M dwarfs (solid black and red dashed lines, respectively). As flares were measured with the bulk photometric variability removed from the lightcurve, the trend between range and EW$_{phot}$ indicates that flares with larger relative energy (as measured by EW$_{phot}$) do take place on stars with greater quiescent variability. The larger photometric variability is likely caused by larger spot coverage$-$ this may in turn indicate either that spottier stars have intrinsically more energy released in a single flare, or perhaps that a greater number of spots implies a greater number of magnetic loops on the stellar surface, increasing the likelihood that multiple loops will be affected by a reconnection event (as in the case of flaring arcades on the Sun). It is notable that the distribution of ranges in our flare sample is fairly typical of other spot$-$modulated stars in the \textit{Kepler} data \citep[e.g.][]{basri2010a}$-$ these stars do not represent a particularly variable sample amongst the rest of the stars in the Quarter 1 observations. Indeed, three stars in the flare sample did not have obvious spot modulation in quiescence. Due to the 30 day duration of Q1 we cannot discern rotation periods of longer than 30 days in the data, so it is possible that these flares are on stars with longer periods that are not evident in Q1 (as opposed to stars that are intrinsically flat in quiescence). We will revisit these stars in future quarters. 

\begin{figure}
\begin{center}
\includegraphics[width=0.95\textwidth]{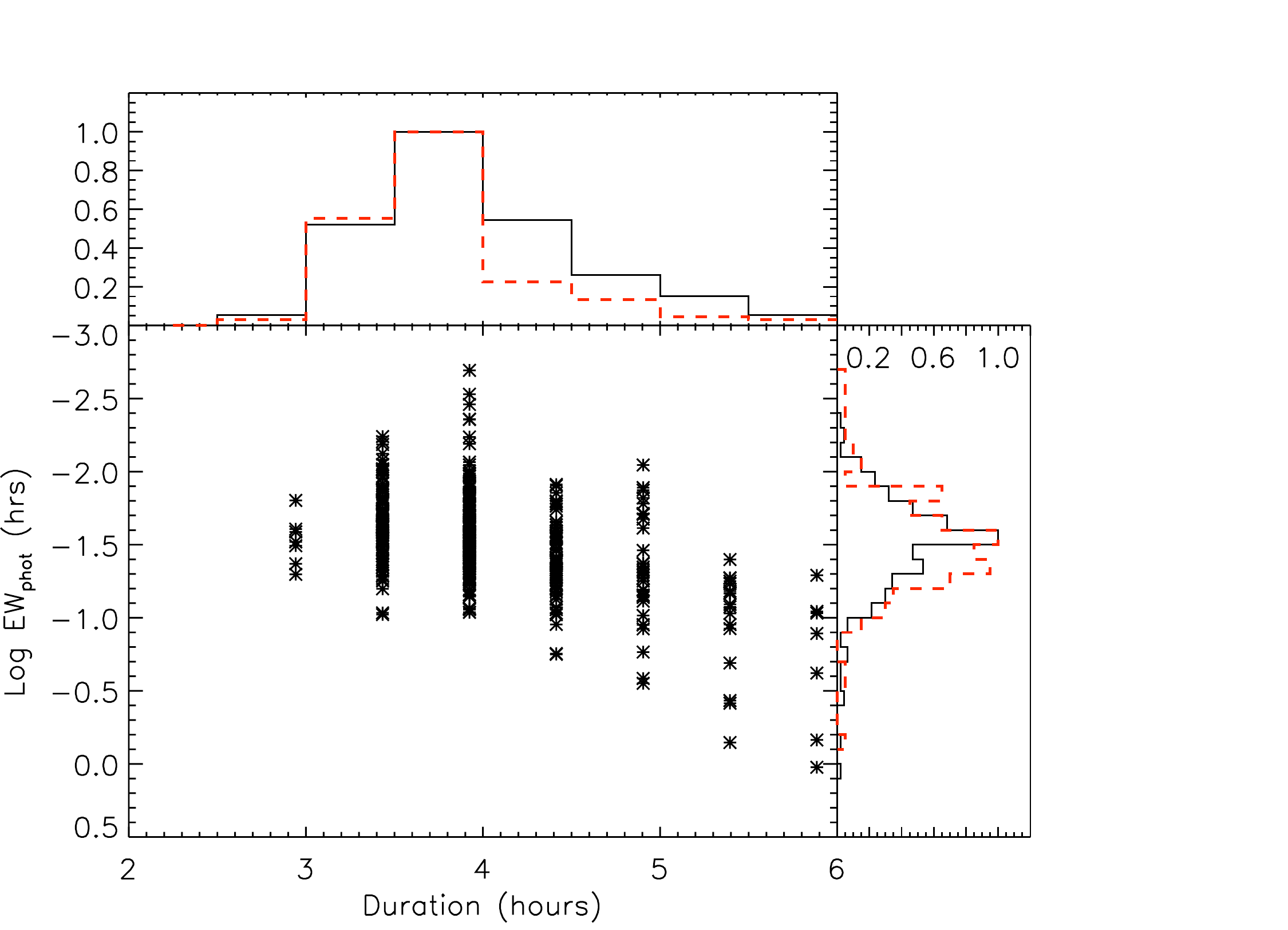}
\end{center}
\caption{Here we show the median duration of flares versus the photometric equivalent width in the central panel. It is evident that longer duration flares tend to have a higher photometric equivalent width, which is to be expected as photometric equivalent width is a time-integrated quantity. In the top and right histogram panels, we show the duration and photometric equivalent widths separated into M dwarfs (red dashed line) and K dwarfs (black solid line). While the M dwarfs tend to have shorter duration flares, they also tend to have higher photometric equivalent widths than the K dwarfs, implying that they release relatively more energy in a given amount of time.}
\label{flare_histnrgdur}
\end{figure}

The duration and frequency of flares is particularly important to understand, both in the context of understanding the effect of flares on planetary habitability, and as a way of estimating Galactic transient rates in time domain surveys. In Figure \ref{flare_histnrgdur}, we show the median duration of all flares on each star versus EW$_{phot}$ for all stars in the main panel, and as histograms of the duration in hours and the photometric equivalent width (at the top and right, respectively) for the K dwarfs (black solid line) and M dwarfs (red dashed line). EW$_{phot}$ increases with the duration of the flares, which is to be expected as EW$_{phot}$ is a time$-$integrated quantity. Interestingly, although the K dwarfs have more flares with durations of more than 4 hours than the M dwarfs, the photometric equivalent widths of the M dwarfs are slightly higher than the those of the K dwarfs. In other words, the M dwarf flares tend to release more energy relative to their quiescent flux in a given amount of time during a flare.

\begin{figure}
\begin{center}
\includegraphics[width=0.9\textwidth]{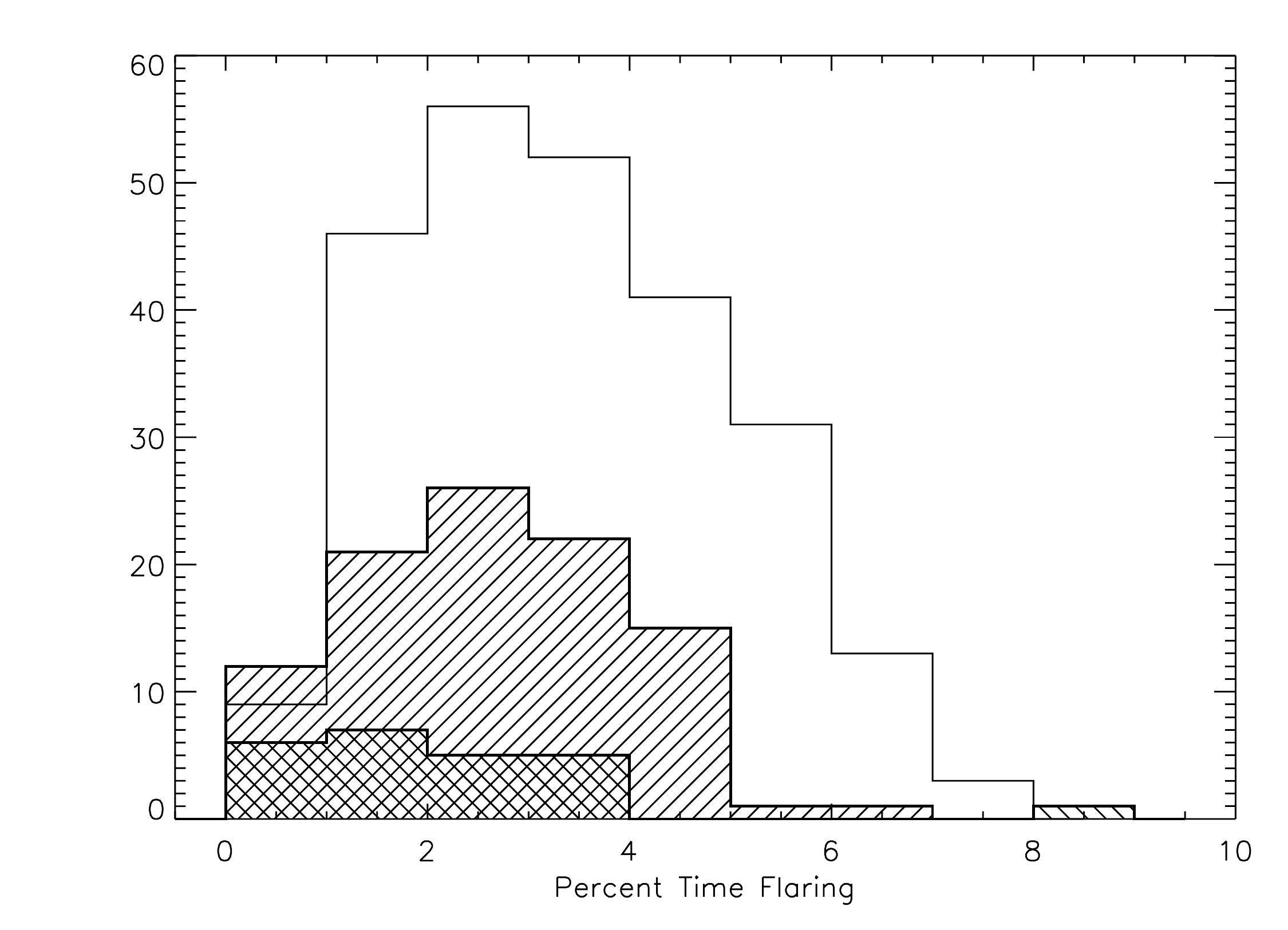}
\end{center}
\caption{In this figure we show histograms of the percentage of time spent flaring binned by the median flare duration (cross hatched: 5 to 6 hour median duration, diagonal filled: 4 to 5 hour median duration, no fill: median duration less than 4 hours.). The stars which have the longest median flares tend to spend the least amount of time flaring overall, implying that some stars may release the majority of the energy in less frequent long duration events, while others flare more frequently for shorter amounts of time.}
\label{flare_perdur}
\end{figure}

We also calculated the flare rates for our sample (defined as the number of flares detected divided by the time interval of the observation, 33.5 days), and the percentage of time spent flaring (defined as the total duration of all flares over the observing time, again 33.5 days). Figure \ref{flare_perdur} shows histograms of the percentage time spent flaring for flares of three bins in duration: 5 to 6 hours' duration (crosshatch filled), 4 to 5 hours' duration (diagonal line filled) and less than 4 hours duration (unfilled). It is evident that stars with longer median duration flares actually spend a lower amount of the overall observation time flaring. In part, this behavior supports previous observations of individual stars where smaller flares were found to be more common than large flares \citep[e.g.][]{lme1976}, but it also indicates that some stars may release the majority of their flare energy all at once in large events, while others undergo more frequent, smaller events. We investigated the magnitude distribution of the stars in each bin to ensure that long duration flares weren't preferentially being detected on brighter stars, and found that the observed effect is not a result of magnitude (and thus detection) bias.  

\begin{figure}
\begin{center}
\includegraphics[width=0.9\textwidth]{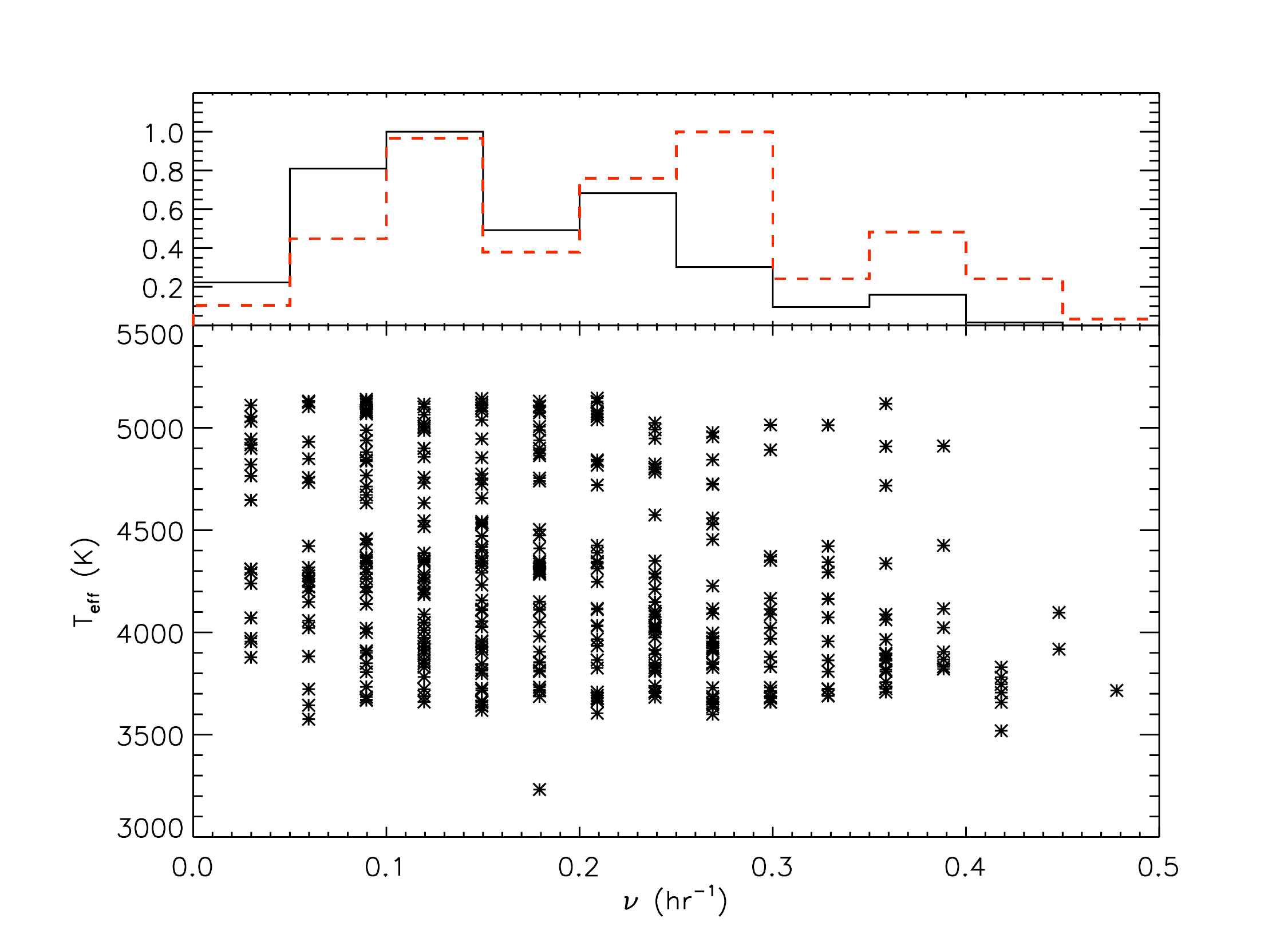}
\end{center}
\caption{In this figure we plot the flare frequency as a function of temperature in the lower panel, with normalized histograms for the flare frequency of the M dwarfs  (red dashed line) and the K dwarfs (black solid line) plotted above. While the two samples have roughly the same amount of stars that flare in the 0.1-0.25 hr$^{-1}$ range, the M dwarfs dominate the population that flares most frequently, while the K dwarfs dominate those who flare least frequently.}
\label{flare_freqvteff_withhist}
\end{figure}

In the lower panel of Figure \ref{flare_freqvteff_withhist}, we show flare rate as a function of effective temperature. The top panel of this figure shows two normalized histograms, one for the K dwarfs (solid black line) and one for the M dwarfs (dashed red line). While the two samples have comparable relative populations of stars with low flare frequency (0.1 $-$ 0.2 hr$^{-1}$), proportionally more M dwarfs make up the most frequently flaring stars, while the K dwarfs dominate the population who flare least frequently. M dwarfs therefore flare more often than K dwarfs, but for shorter durations, and during flares release more energy relative to their quiescent flux in a given time than their more mssive cousins.

\begin{figure}
\begin{center}
\includegraphics[width=0.95\textwidth]{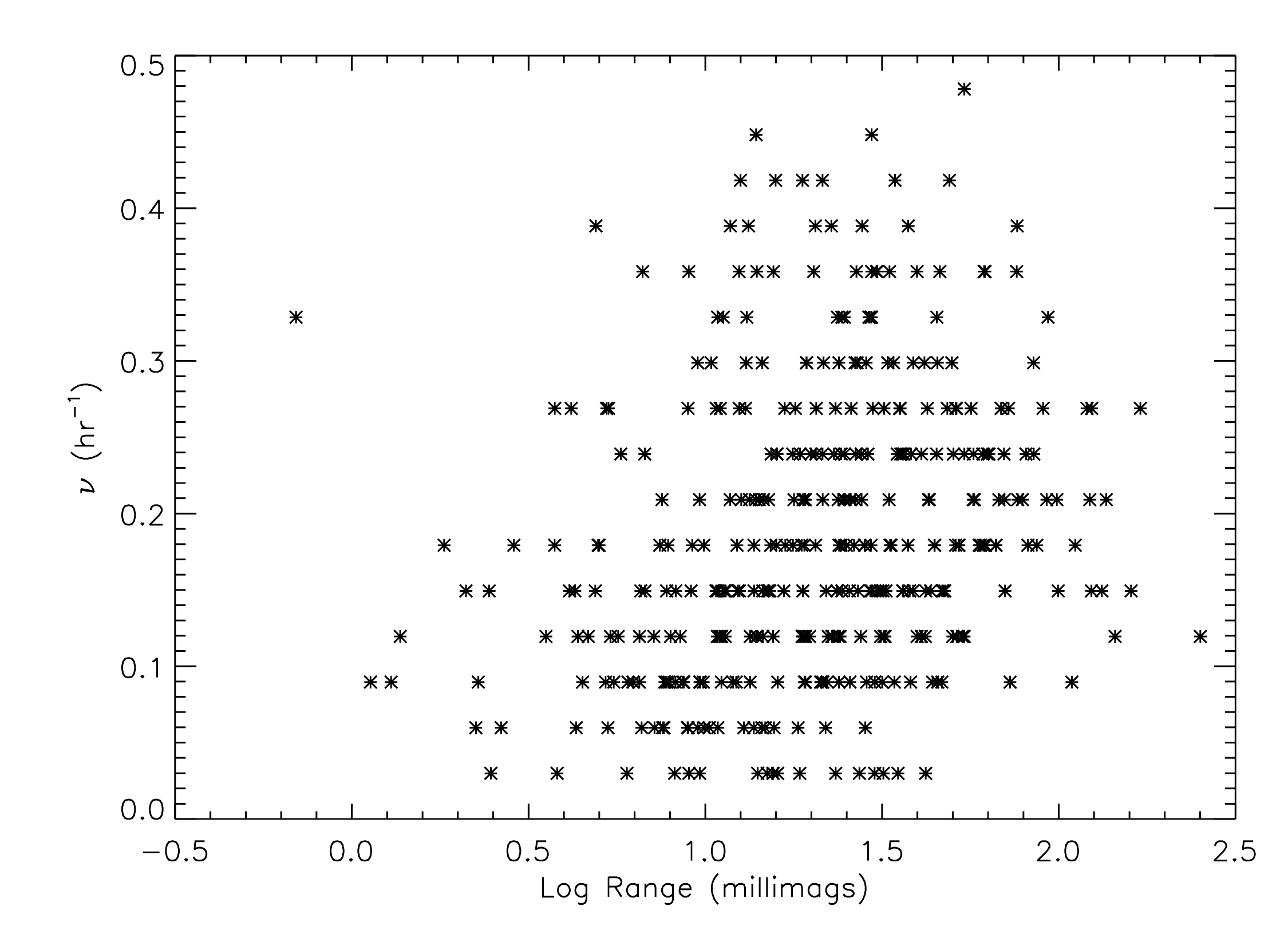}
\end{center}
\caption{Here we show the distribution of flare frequency with the variability range. There is a wide spread of range with flare frequency, though it does seem that stars that are somewhat more variable in quiescence tend (weakly) to flare more often.}
\label{flare_rangevfreq}
\end{figure}

In Figure \ref{flare_rangevfreq}, we show that while there is a loose correlation between the photometric range of our sample and the flare frequency, overall stars of different flare frequency are found over a large spread in range. 
One might expect that larger photometric variability (presumably due to larger spot coverage) might lead to more frequent flaring$-$the greater the number of spots, the greater the number of magnetic loops protruding from the stellar surface, and therefore the more potential sites for flares to take place. The fact that this is only a weak correlation implies that the relationship between having a great number of magnetic structures and the frequency of flares is not strictly causal$-$ if a flare event in one active region played some role in ``triggering'' flares in neighboring regions, one would likely see a more direct correlation between the range and the flare frequency. 

\begin{figure}
\begin{center}
\includegraphics[width=0.95\textwidth]{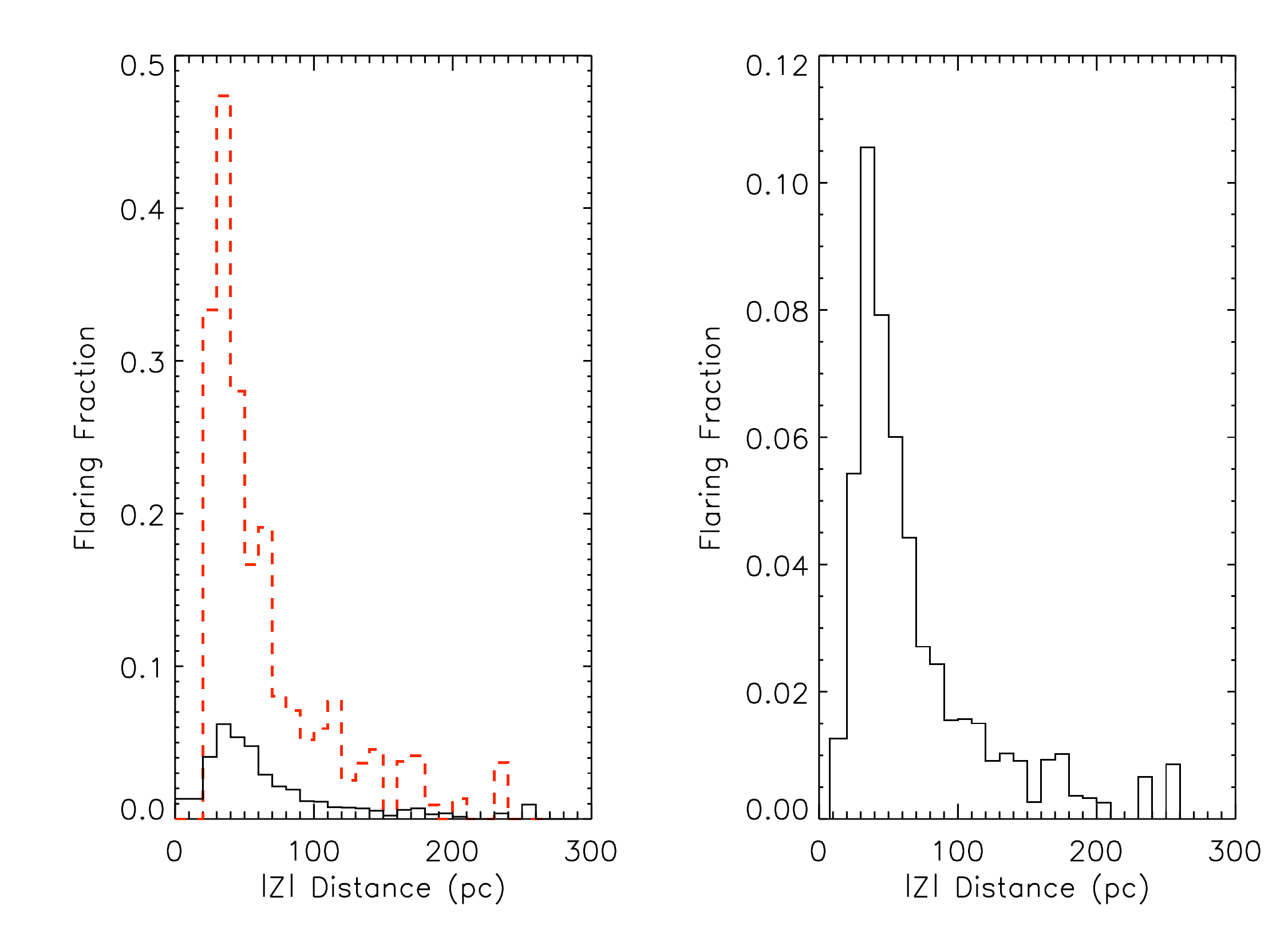}
\end{center}
\caption{At left we show the flaring fraction of the M dwarfs (red dashed line) and K dwarfs (solid black line) as a function of the distance above the Galactic plane, while at right we show the flare fraction as a function of distance for the entire flare sample.}
\label{flare_flrfracdist}
\end{figure}

Lastly, we examined the spatial distribution of the flare sample, as it has been noted that cool stars with a greater $\vert$Z$\vert$ distance from the Galactic plane flare less frequently than stars in the solar neighborhood \citep{west2008}. This observed drop$-$off in flare rate is due to the fact that older, less active stars have had a longer time in which to undergo dynamical interactions, thus scattering them further out of the plane. As \textit{Kepler} observes in only one field (100 deg$^2$ centered on 19h 22m 40s, $+$44$^{\deg}$ 30$\arcmin$ 00$\arcsec$), an estimate of the $\vert$Z$\vert$ distance is relatively simple trigonometry. We take the Sun to be 15 pc above the plane (Cohen 1995, Binney et al. 1997, Ng et al. 1997). All the stars in our flare sample have SDSS and 2MASS colors, for which there are existing photometric parallax relations \citep{bochanski2010,covey2007}. The left panel of Figure \ref{flare_flrfracdist} shows the distribution of the flaring fraction for the entire sample as a function of distance; clearly stars in the local neighborhood make up the majority of the flare sample.  The right panel of this same figure shows the flare fraction by spectral type (K dwarfs: solid black line, M dwarfs: dashed red line)$-$ the local flare fraction is heavily dominated by the M dwarfs.  This result fits nicely with previous studies of active M dwarfs, e.g. \citet{hgr1996}, who found that up to $\sim$60$\%$ of M dwarfs in the solar neighborhood had H$\alpha$ in emission.

\section{Conclusions and Future Work}

The \textit{Kepler} data provide a new view of white light stellar flares, sampling a population of stars of varying activity rarely included in traditional flare studies. \textit{Kepler}'s unprecedented photometric precision also provides simultaneous information on the quiescent variability of these flare stars. In this work, we have defined a new flare measure, the photometric equivalent width (EW$_{phot}$), which expresses the flare energy relative to the quiescent luminosity of the star. We conducted an automated search for flares amongst the cool dwarfs (classified in the \textit{Kepler} Input Catalog  as T$_{eff}$ $\le$ 5150 K and logg $\ge$ 4.2) during the 33.5 days of \textit{Kepler}'s Quarter 1 observations. We vetted the candidate events by eye, and found 373 stars which possessed one or more obvious white light flares. 

In comparing the photometric variability of the quiescent lightcurves (as quantified by our ``variability range'' statistic), we observe that stars with larger quiescent variability (and thus larger spot coverage) seem to have intrinsically larger flares relative to their quiescent energy output (i.e. larger EW$_{phot}$). We also find that the distribution of observed flare peaks as a function of effective temperature is somewhat flat$-$$-$ in part this trend is caused by the difficulty in detecting smaller flares on the warmer K dwarfs, as the contrast between the white light flare emission and the stellar photosphere is less than for cooler stars. However, we find fewer flares with large peaks on the M dwarfs versus the K dwarfs, while the M dwarfs tend to have somewhat larger EW$_{phot}$ and shorter duration flares than the K dwarfs. Taking these observations together, we infer that M dwarf flares evolve faster than K dwarf flares, such that their flaring energy output relative to their quiescent luminosity is higher than that of the K dwarfs.

The time domain behavior of our flare sample is also quite interesting$-$ in examining the frequency of flares, binned by the median duration of flares on each star, we find that longer duration flares tend to happen on stars that flare less frequently. Some stars may therefore release the majority of their flare energy in less frequent long duration events, while others release relatively less energy in a single event but flare more often. This trend appears to be a function of spectral type: overall, the M dwarfs make up the most frequently flaring stars in our sample, flaring more often (but for shorter durations) than K dwarfs, and release more energy relative to their quiescent flux in a given time than their more massive K dwarf cousins. While a single month of data is too short to properly comment on the probability distribution of flares, the ongoing monitoring of these flare stars with \textit{Kepler} will allow us to delve into that question in the near future.

Our flare sample is distributed over a few hundred parsecs above the Galactic plane, allowing us to calculate the flare fraction of our stars as a function of $|$Z$|$. In agreement with previous studies of M dwarf flare activity, we find that the flaring fraction of M dwarfs reaches roughly 50$\%$ in the local neighborhood. The K dwarfs have a far lower flare fraction, under 10$\%$ even locally, but as there are so many more K dwarfs in our sample, their population effectively dilutes the flare fraction when the sample is taken as a whole, such that only $\sim$10$\%$ of nearby stars flare. 

This work provides important clues to the ongoing question of how stellar flares affect planetary habitability. The frequency of flares is a crucial piece of information because after a flare impacts and modifies the photochemistry of the planetary atmosphere, the atmosphere takes some time to return to its preflare state. If flares recur on shorter timescales than the atmospheric re$-$equilibration time, it is possible that the buildup of photochemical flare by$-$products will permanently alter the planet's atmosphere (e.g. Segura et al. 2010). To the many open questions regarding planetary habitability, we add one more: in the establishment or maintenance of a hospitable planet, is it preferable to be affected by longer duration, less frequent flares, or shorter, more frequent and more energetic flares? This question is most relevant during the youth of the star-planet system-- activity in cool stars, M dwarfs in particular, declines strongly with age \citep{west2008}. The distribution in $|$Z$|$ of our sample indicates that we are preferentially studying a young thin disk population, which flares relatively frequently-- the majority of these stars' exceedingly long main sequence lifetimes are spent in relative quietude.

In future work, we intend to compare the rate and relative energy of flares with stellar rotation period, using the rotation analysis discussed in \citet{basri2010b}. Additional quarters of \textit{Kepler} data will provide longer duration lightcurves, permitting us to associate flare properties with rotation rate and differential rotation.  

We would also like to revisit smaller flares in the sample. The \textit{Kepler} pixel-level data (the actual pixels in each aperture, rather than their summed total) will eventually become available for these stars. These data can be used to ascertain the astrophysical nature of relatively small brightness changes by seeing whether the change in brightness is associated with the target star or is caused by a brightness fluctuation on a neighboring pixel. We also intend to address the white light flare occurrence on solar type stars$-$ although the contrast of the white light emission against a roughly solar temperature photosphere is expected to be quite small, \textit{Kepler}'s precision offers the best chance of detecting such events. While the white light emission on the Sun is small and quite localized, it is as yet unknown whether that is a trait of solar type stars in general, or just one solar type star in particular.  

\acknowledgements
LMW is grateful for the support of the \textit{Kepler} Fellowship for the Study of Planet$-$Bearing Stars. Funding for this Discovery mission is provided by NASA's Science Mission Directorate.

\clearpage

\begin{figure}
\begin{center}
\includegraphics[width=0.95\textwidth]{f6}
\end{center}
\caption{In the center panel we show the variability range versus the photometric equivalent width for the M dwarfs (red diamonds) and the K dwarfs (black asterisks). At top and right we show normalized histograms of these quantities, with the M dwarfs shown as a red dashed line and the K dwarfs as a solid black line. The two samples are roughly comparable in range, with the K dwarfs being slightly more variable in quiescence than the M dwarfs, while the M dwarfs have a somewhat higher photometric equivalent width than the K dwarfs. }
\label{flare_histrangevenergy}
\end{figure}

\clearpage
 commands



\end{deluxetable}

\end{document}